\documentstyle[preprint,prb,aps]{revtex}

\begin{document}

\draft

\title{
Variational Hilbert space truncation approach to quantum Heisenberg
antiferromagnets on frustrated clusters
}
\author{ N. A. Modine and Efthimios Kaxiras}
\address{
Department of Physics, Harvard University, Cambridge MA 02138
}
\date{\today}

\maketitle

\begin{abstract}
We study the spin-$\frac{1}{2}$ Heisenberg antiferromagnet on a
series of finite-size clusters with features inspired  by the
fullerenes.  Frustration due to the presence of pentagonal rings
makes such structures challenging in the context of quantum Monte-Carlo
methods.  We use an exact diagonalization approach combined with a
truncation method in which only the most important basis states
of the Hilbert space are retained.  We describe an efficient variational
method for finding an optimal truncation of a given size which
minimizes the error in the ground state energy.  Ground state energies
and spin-spin correlations are obtained for clusters with up to
thirty-two sites without the need to restrict the symmetry of the
structures.  The results are compared to full-space calculations
and to unfrustrated structures based on the honeycomb lattice.

\end{abstract}

\pacs{ PACS 75.10.Jm}

\section{Introduction}
\label{Introduction}

The spin-$\frac{1}{2}$ Heisenberg antiferromagnet (HAFM) has long been
studied as a simple example of a strongly interacting quantum many-body
system~\cite{manousakis91}.  Recently, it has attracted considerable attention
in the context of the copper oxide high-temperature
superconductors~\cite{anderson87,zhang88}.  The Hamiltonian of the HAFM
is given by
\begin{equation}
   H=J\sum_{\langle i,j \rangle}\vec{S}_{i}\cdot\vec{S}_{j}
      \equiv J\sum_{\langle i,j \rangle} S_{i}^{z}S_{j}^{z} +
      \frac{1}{2}\left(S_{i}^{+}S_{j}^{-}+S_{i}^{-}S_{j}^{+}\right)
\end{equation}
where $J$ takes positive values, $\langle i,j \rangle$ refers to nearest
neighbor pairs, $\vec{S}_{i}$ is the spin operator for a spin-$\frac{1}{2}$
located at site $i$, and $S_{i}^{+}$ and $S_{i}^{-}$ are the corresponding
raising and lowering operators.  The operator
$S_{i}^{+}S_{j}^{-}+S_{i}^{-}S_{j}^{+}$ exchanges antiparallel spins,
but vanishes when applied to a pair of parallel spins.  The terms of
this type produce off-diagonal matrix elements equal to
$\frac{J}{2}$ between basis states (i.e.\ spin configurations) that are
related by a single exchange of nearest neighbor spins.  The terms of
the form $S_{i}^{z}S_{j}^{z}$ combine to give a diagonal matrix element
for each state equal to $\frac{J}{4}$ times the difference between the
number of parallel nearest neighbor spins and the number of anti-parallel
nearest neighbor spins in that configuration.  Despite the simplicity
of the model, no analytic solutions have been found for nontrivial
structures except in one dimension~\cite{manousakis91}.

Since the Hamiltonian is invariant under uniform rotations of the spins,
one can choose its eigenstates to be simultaneous eigenstates of the
operators $\vec{S}_{TOT}^{2}$ and $S_{TOT}^{z}$, where $\vec{S}_{TOT}$
is the total spin.  For a system containing an even number of spins $n$,
whatever the ground state value of $\vec{S}_{TOT}^{2}$, there is always
a ground state with $S_{TOT}^{z} = 0$.  Therefore, a ground state can
always be found in the subspace spanned by the
\begin{equation}
   N_{total} = \frac{n!}{(n/2)!(n/2)!}
\end{equation}
basis states with an equal number of up and down spins.
The generalization to an odd number of spins is straightforward.  Since
the Hamiltonian is real, the ground state eigenvector can be chosen to
be real.

   In this paper, we solve this model for a series of structures
that embody the basic structural features of the fullerenes, which are
spherical shells of threefold coordinated carbon atoms arranged
in pentagonal and hexagonal rings.  It can be shown that every such
structure must have twelve pentagonal faces~\cite{growth_and_form}.
The total number of sites can be varied by changing the number of hexagons.
The smallest such structure contains no hexagons and has $20$ sites.
Figure \ref{frustrated_structures} shows several fullerene related
structures that we discuss in this paper.  We shall refer to the
structures in Fig.~\ref{frustrated_structures}~(a)--(e) as F-20, F-24,
F-26, F-28, and F-32, respectively.  For simplicity, we shall treat
all of the bonds in these structures as equivalent even though in
actual carbon clusters they may differ.
On a pentagonal ring, it is impossible to arrange
all spins in an antiferromagnetic pattern.  This introduces frustration
in the classical ground state where nearest neighbor spins would prefer
to be antiparallel.  For comparison, we also study several unfrustrated
structures that are derived from the honeycomb lattice by applying
periodic boundary conditions.  These structures are shown in
Fig.~\ref{hexagonal_structures}~(a)--(c).  We refer to these structures
as H-18, H-24, and H-26, respectively.  These structures have toroidal
topology rather than the spherical topology of the frustrated structures.
Table \ref{struct_props} summarizes the geometrical features of the
structures that we investigate.

   A group of powerful techniques used to investigate quantum many-body
systems such as the HAFM are based on quantum Monte-Carlo methods.
In systems with frustration, these methods either require the summation
of a very large number of terms with alternating signs (known as the sign
problem) or depend on a ``guiding'' wavefunction which must be properly
guessed.  Here we use a different approach based on exact diagonalization
of the Hamiltonian matrix.  This approach has the advantage of not being
affected by the sign problem, but is limited to rather small system sizes
because the number of states in the Hilbert space grows exponentially with
the size of the
system.  For example, in Table \ref{struct_props} we list the number of
states in the $S_{TOT}^{z} = 0$ subspace for each cluster that we
investigate.  Thus, it takes a major increase in either computer power
or efficiency of the algorithm to get a modest increase in the size of
system that can be investigated.

Using exact diagonalization techniques, modern computers can handle
systems with $\leq 36$ spins.  A $36$ spin system has about $9$
billion basis states in the subspace with $S_{TOT}^{z} = 0$.
The Hamiltonian matrix is
sparse and has only about $300$ billion nonzero entries for
this size system.  Memory constraints make it difficult to store this
matrix.  The symmetries of the structure must be used to reduce the size
of the basis space in order to make calculations tractable.  The usefulness
of symmetrization depends on how many mutually commuting symmetry operations
can be found.  Symmetry is most useful for lattices where all translations
commute, such as the square lattice.  Even noncommuting symmetries could
be easily exploited if the ground state was known to transform
according to the identity representation of the symmetry group.  This can
not be assumed to be the case for the frustrated HAFM.  To our knowledge,
the largest structure that has been solved using exact diagonalization
and taking advantage of all of its symmetries is the 36-site square
lattice~\cite{schulz92}.  It would be very difficult to find the
ground state of a structure with the same size and a lower number
of commuting symmetries without approximation.

   One way to manage larger systems is to restrict the wavefunction to
the space spanned by a subset of the basis states.  In this approach,
the problem is transformed into finding a subspace that accurately
approximates the full-space result, but that is small enough to be handled
computationally.  In this paper, we variationally optimize the truncation
of the Hilbert space, and exactly diagonalize within the truncated space.
The rest of the paper is organized as
follows: section \ref{Choice of Truncation} contains a justification
for our choice of optimal wavefunction and truncated space, section
\ref{Results} discusses the ground state properties that we obtain
with this approach for a series of frustrated and honeycomb clusters,
and section \ref{Conclusion} summarizes our conclusions.

\section{Choice of Truncation}
\label{Choice of Truncation}

Consider a truncation of the space to the basis states
$\left\{|\alpha_{1}\rangle,|\alpha_{2}\rangle,\ldots,
|\alpha_{N_{trunc}}\rangle\right\}$
where $N_{trunc} < N_{total}$.  Define a truncated Hamiltonian that
consists of those elements of the original Hamiltonian that connect
states retained in the truncated space.
Let $E(\{\alpha_{1},\alpha_{2},\ldots,\alpha_{N_{trunc}}\})$ denote
the smallest eigenvalue of the truncated Hamiltonian.  We define the
optimal truncation as the one that minimizes $E$ with respect to all
sets with $N_{trunc}$ basis states.  By the variational principle,
the ground state wavefunction of the corresponding truncated Hamiltonian
is the wavefunction that, subject to the constraint of vanishing for all
but $N_{trunc}$ states, minimizes the expectation of the full Hamiltonian.
Therefore, $E$ for the optimal truncation is the smallest possible
variational upper bound on the true ground state energy that can be
obtained using trial wavefunctions that have no more than $N_{trunc}$
nonzero components.

The minimization over sets of basis states is accomplished
using a stochastic search:  An initial truncation is chosen and the ground
state energy of the corresponding truncated Hamiltonian is found using
the Lanczos method~\cite{sorensen92}.  Moves in the stochastic search consist
of adding states to the space and eliminating others while keeping the
overall number of states fixed.  The Lanczos method is used at each step to
find the ground state energy for the new truncation, and the move is accepted
or rejected according to a Metropolis
algorithm~\cite{metropolis53,numerical_recipes}.
This procedure is repeated until all new moves are rejected, in which case
a minimum of $E(\{\alpha_{1},\alpha_{2},\ldots,\alpha_{N_{trunc}}\})$ has
been found.  If this is the global minimum, the resulting truncation is
the ideal truncation.  We have found no evidence that the procedure gets
trapped in local minima.

For systems that are small enough that the full problem can be solved, we
have also applied an alternative truncation procedure for purposes of
comparison with our variational scheme.  This consists of keeping only
the basis states that have the largest weights in the full-space ground state
solution and varying the cutoff weight below which states are excluded from
the basis.  The energy obtained from this alternative procedure must be
greater than or equal to the variational result, but the wavefunction
from this alternative procedure is expected to be closer to the true ground
state.  Therefore, this alternative procedure might be expected to yield
better results for correlation functions.  A comparison of results obtained
using these two independent methods helps to assure that the variational
procedure is converging properly and shows that the procedure produces
reasonable correlation functions.

  In order to optimize the variational search, it is necessary to bias the
selection of the states to be added to, or eliminated from, the truncated
basis during each step.  The procedure proposed here is analogous to
force-bias Monte Carlo.  In our case, the equivalent of the force in
a particular direction is the difference between the energy when a
particular state $|\beta\rangle$ is included in a truncation
and the energy when the state is not included in the truncation:
\begin{equation}
    \nabla_{\beta}E(\{\alpha_{1},\alpha_{2},\ldots,\alpha_{N_{trunc}}\})
        \equiv E(\{\alpha_{1},\alpha_{2},\ldots,\alpha_{N_{trunc}},\beta\})
         - E(\{\alpha_{1},\alpha_{2},\ldots,\alpha_{N_{trunc}}\}).
\end{equation}
In force-bias Monte Carlo, the force is a function of the configuration
of the system, and correspondingly $\nabla_{\beta}E$ is a function of
the set of states included in the truncation.  In our case, since each
state is either included or not included, we must take
\begin{equation}
\nabla_{\beta}E(\{\alpha_{1},\alpha_{2},\ldots,\alpha_{N_{trunc}},\beta\})
   \equiv \nabla_{\beta}E(\{\alpha_{1},\alpha_{2},\ldots,\alpha_{N_{trunc}}\}).
\end{equation}
$\nabla_{\beta}E$ can be estimated easily for each $\beta$ using the solution
from the previous truncation:  We denote the states that are included in the
previous truncation as internal states, and the remaining states of the
full Hilbert space as external states.  The internal states are the states
that could be eliminated from the previous truncation in the process of
forming the new truncation, while the external states are the states that
could be added.  For each internal state, we wish to calculate the change
in the variational energy caused by eliminating it from the previous
truncation.  Let the ground state wavefunction for the previous truncation be
$|\Psi_{0}\rangle$ and let $\psi_{\beta} = \langle\beta|\Psi_{0}\rangle$.
We approximate the ground state of the truncation with the state
$|\beta\rangle$ \underline{eliminated} by assuming that the rest of the
wavefunction remains unchanged except for an overall normalization factor,
\begin{equation}
   |\Psi_{0-\beta}\rangle =
       \frac{|\Psi_{0}\rangle - \psi_{\beta}|\beta\rangle}
            {\sqrt{1-{|\psi_{\beta}|}^{2}}}.
\end{equation}
To first order in ${|\psi_{\beta}|}^{2}$ this approximation gives,
\begin{equation}
      \nabla_{\beta}E =
         E_{0} - \langle\Psi_{0-\beta}|H|\Psi_{0-\beta}\rangle =
         {|\psi_{\beta}|}^{2}\left(E_{0}-H_{\beta\beta}\right)
\label{inside_force}
\end{equation}
where $H|\Psi_{0}\rangle = E_{0}|\Psi_{0}\rangle$ and
$H_{\beta\beta} = \langle\beta|H|\beta\rangle$.
Similarly, the effect of \underline{adding} an external state is
approximated using second order perturbation theory as:
\begin{equation}
    \nabla_{\beta}E =  \frac {\left|\langle\beta|H|\Psi_{0}\rangle\right|^{2}}
                        {E_{0}-H_{\beta\beta}}.
\label{outside_force}
\end{equation}
Note that $\nabla_{\beta}E$ will be zero if $|\beta\rangle$ is neither
an internal state nor an external state that is connected by the
Hamiltonian to an internal state.  Depending on the stage of the
variational procedure, a set of trial states is chosen which either
contains all of the states for which $\nabla_{\beta}E$ is nonzero,
or a randomly chosen subset of such states.  $\nabla_{\beta}E$ is
calculated for this set, and the new truncation is formed by taking
the states with the largest values.  Choosing a random subset of
trial states introduces a stochastic element into the computation
and effectively reduces the variational step size.  During a
minimization procedure where the full set of trial states is used
at every step, a move will eventually be rejected in the Monte-Carlo
evaluation.  Further iterations beyond this point will simply generate
the same move.  This is similar to a gradient minimization with a
fixed step size where the step overshoots the minimum.  Here, since
each state is either included or not included, it is impossible to
reduce the step size in the usual sense.  Instead, the step size can
be effectively reduced by using a randomly chosen subset of
the components of the gradient.  The fastest minimization is
achieved by using all of the trial states until the first move is
rejected, and then considering a random subset which is gradually
reduced in size.  For the HAFM model considered in this paper,
we found that our move selection algorithm was so effective that
additional moves after the first rejected move produced minimal
improvements in the energy.  Accordingly, we stop the variational
procedure when the first move is rejected.

The idea of iterative improvement of a Hilbert space truncation
using perturbative estimates of the importance of new states has a
long history in the quantum chemistry
literature~\cite{bender69,huron73,evangelisti83,feller89,harrison91}.
In addition, for this class of problems,
the final truncated results are typically
corrected with a perturbative treatment of the remaining
states~\cite{maynau91,shavitt92,wenzel92,steiner94}.
Extrapolation methods are also frequently used~\cite{buenker75}.
Such methods would likely be a useful addition to our method,
but since the emphasis of this paper is on a variational approach,
we have avoided such corrections.
Iterative improvement of a Hilbert space truncation has also been studied
in the context of quantum lattice models.  De Raedt and von der Linden
estimated the importance of a new basis state by means of the energy lowering
obtained from a Jacobi rotation involving the state~\cite{de_raedt92}.
Riera and Dagotto added basis states that are connected by the Hamiltonian
to states with a large weight in the current truncated solution~\cite{riera93}.
In this previous work, the basis is expanded by adding selected new
states until either the desired quantities converge or computational
limits are reached.  In contrast, our emphasis is on finding the optimal
basis of a given size.  Working with a constant size basis has two
advantages:

(1) It allows us to define the optimal basis in an unambiguous
manner and to express the problem of finding this optimal basis
as a minimization problem.
This makes it possible to harness the full power of the Metropolis
algorithm and the simulated annealing approach.

(2) It allows us to tackle problems with no clear hierarchy of importance
among the basis states.  In quantum chemistry, there is a hierarchy of
states in which higher excitations are progressively less important.
In contrast,
the frustrated HAFM lacks any clear {\em a priori} hierarchy
among the basis states.  As a result, truncation can
induce level crossings and change the character (e.g. the symmetry)
of the ground state.  If a basis selection process were to start with
an incorrect ground state, augmentation of the truncation runs
the risk of not selecting the basis states that are
important for the true ground state. This makes it likely that the true
ground state would never be found.  By working with a basis of a
constant size, which is variationally optimized, we avoid this problem.

The effectiveness of the variational Hilbert space truncation procedure
can be demonstrated by comparing its results to those obtained from
the full-space solution.  Define the fractional error in the energy for
a given truncation by
\begin{equation}
   \delta\epsilon(\{\alpha_{1},\alpha_{2},\ldots,\alpha_{N_{trunc}}\})=
    \frac{E(\{\alpha_{1},\alpha_{2},\ldots,\alpha_{N_{trunc}}\})-E_{N_{total}}}
          {E_{N_{total}}},
   \label{def_of_dE}
\end{equation}
where $E_{N_{total}}$ is the full-space ground state energy.
Figure \ref{energy_errors} shows $\delta\epsilon$ for the truncation resulting
from the variational truncation procedure and the truncation resulting from
keeping the states with the largest weights in the full-space solution.
The energies found using the variational procedure are just slightly below
those found by truncating based on the full-space solution.
The fact that the variational energies are the lowest energies indicates
that the variational minimization is converging properly.  The closeness
of the two results indicates that our definition of a best truncation
is successful in capturing the most important
parts of the full-space wavefunction.  The difference between the two
results grows as the retained fraction of the space diminishes and as
the physical system gets smaller, but it stays relatively insignificant
except for the smallest truncation size of the smallest structure.  For
example, retaining only $\frac{1}{6}$ of the basis states of the F-20
structure results in only about 1 percent error in the energy.  Note
that in order to get the same fractional error, a smaller fraction of
the basis vectors is required for the larger systems.  As a result,
the number of states that must be retained in the truncated space grows
more slowly than the number of states in the full-space.  Therefore,
larger systems make truncation increasingly useful.  The curves resulting
from the frustrated structures have a different shape than the curves
resulting from the unfrustrated structures.  The error falls more slowly
for the unfrustrated structures than for the frustrated structures as the
retained fraction of space increases.  This suggests that the method
is more useful for frustrated structures.

Figure \ref{best_case_corr} shows the correlations for the honeycomb lattice
structures as a function of the fraction of space retained in the truncation.
Since for these structures the nearest neighbor correlation function is
proportional to the energy, it is not included.  The multiple lines
are due to the fact that the 24 and 26 site structures each have two
inequivalent 3rd neighbor correlations, and the 24 site structure has two
inequivalent 4th neighbor correlations.  Again, both the results of the
variational truncation method and the results of truncating the Hilbert
space based on the weights of states in the full-space solution are shown.
The truncation based on the full-space solution is
expected to give a better approximation to correlation functions than
the variational method, but for the correlations considered here,
the results of two methods are
almost indistinguishable.  Furthermore, truncation down to a few percent
of the space by either of these methods introduces only a few percent error
in the correlations.  Since the HAFM on the honeycomb lattice has long range
order, all of the correlations are fairly large in magnitude.  This causes
our truncation methods to give particularly good results for these
correlations.

In contrast, correlations between sites that are far apart on the frustrated
structures are a worst case situation.  Correlations on the frustrated
structures usually become very small at long distances.  As a result, the
fractional error in these correlations is quite large.
Figure \ref{worst_case_corr}
shows the fractional error in the correlation that is smallest in magnitude
for the 20, 24, and 26 site frustrated structures.  The full-space values
of these correlations are $3.31 \times 10^{-2}$, $-3.43 \times 10^{-3}$, and
$2.02 \times 10^{-3}$, respectively.  With less than half of the space
retained, the fractional error introduced in these correlations becomes
substantial.  The error resulting from the variational truncation method
is rather similar to the error introduced by truncating the Hilbert space
based on the weights of states in the full-space
solution.  The fractional error in a correlation seems to grow with the
inverse of the magnitude of the correlation.

Since we are interested in the most accurate approximation to the
full-space properties of the system, it is desirable to make
the size of the truncation as large as possible.  As mentioned above,
memory is the primary constraint on the size of the system that can
be handled using exact diagonalization techniques.  Thus, effective
implementation of this algorithm requires careful treatment
of memory usage.  The requirement of maximizing speed while minimizing
memory usage provides a particular programming challenge to
implementing the variational Hilbert space truncation method.
We have implemented the method on the Naval Research Laboratory's
256 node Thinking Machines Corporation CM-5E supercomputer.
In the Appendix we provide an outline of technical issues related to
our implementation of the algorithm on this massively parallel architecture.

\section{Results}
\label{Results}

Table \ref{fs_results} summarizes some of the ground state properties of
the HAFM on the structures we considered.  The expectation of
$\vec{S}_{TOT}^{2}$ can be calculated by summing the correlation
functions between all pairs of sites.  Since each structure
considered has an even number of spins, the possible exact eigenvalues
are $s(s+1)$ where $s$ is an integer.  Deviation from these values can be
expected for truncated solutions because the truncation procedure breaks
the invariance of the model under global spin rotation.  For each of our
full-space solutions (which includes all structures studied except F-32),
the expectation of $\vec{S}_{TOT}^{2}$ is $0$ to the accuracy of the
solution.  Thus, for every system except F-32, the calculated ground
state is a spin singlet.  For the truncated solution of the F-32 system,
this expectation is $\approx 0.5$.  This value is between the values
expected for a spin singlet ($s(s+1) = 0$) and a spin triplet ($s(s+1) = 2$).
It is much closer to the value of the spin singlet than to the triplet.
Moreover, we have found that the variational procedure tends to decrease
this value, indicating that the ground state of F-32 is also a spin singlet.
Table \ref{fs_results} contains two entries for the F-28 structure because
its ground state is a rotational doublet.  The rest of the states
are rotational singlets.  The two F-28 states are distinguished by
considering their transformation properties under improper rotation
about the symmetry axis through the center of the bond between site $19$ and
site $20$ (see Fig.~\ref{frustrated_structures}~(d)).  Under this
transformation, the F-28A state has eigenvalue $1$, while the F-28B state
has eigenvalue $-1$.

The first column of Table \ref{fs_results} contains
the ground state energy per site.  As expected, frustration raises the
ground state energy.  The energies per site of the structures based on
the honeycomb lattice reveal the expected finite size effects for the
HAFM on a lattice: the energy per site increases as the size of
the system increases.  Finite size effects are not as clearly evident in
the frustrated structures, but the trend from F-24 to F-26 to F-28 is
rather similar to what could be expected from finite size effects.
The trend is reversed in F-32.  These clusters are not especially similar
to each other except for overall
topology, so it is reasonable that finite size effects are obscured by
effects due to details of the structure.  Furthermore, as the size of
the frustrated structures increases, the hexagonal rings become more
plentiful and closer together.  Thus, these systems should behave more
like the unfrustrated structures at larger sizes.  Eventually,
the energy must decrease toward the unfrustrated value.  It is likely
that the drop in energy between F-28 and F-32 indicates the beginning
of this trend.  Note that this drop in energy can not be a result of
using a truncated solution for the F-32 system since the energy
resulting from the truncation must be greater than the full-space energy.

The rest of the columns in Table \ref{fs_results} show the nearest
neighbor spin-spin correlations.  The correlation between site
$i$ and site $j$ is defined by
\begin{equation}
   C_{i,j} = \langle\Psi_{0}|\vec{S}_{i}\cdot\vec{S}_{j}|\Psi_{0}\rangle
\end{equation}
where $\Psi_{0}$ is the ground state wavefunction.  The sum of all of
the nearest neighbor correlations for a particular structure
gives the ground state energy. Even though the ground state energies
vary relatively little, the nearest neighbor correlation functions
vary dramatically (see Table \ref{fs_results}).
The nearest neighbor correlations are divided into four columns.  The
column labeled $H-H$ contains correlations between sites that
are both located on the same hexagonal ring.  The column labeled
$H-H^{\prime}$ contains correlations between sites that are
located on two different hexagonal rings.  The column labeled
$H-P$ contains correlations between a site located
on a hexagonal ring and a site that is not located on any hexagonal
ring.  The column labeled $P-P$ contains correlations between
two sites neither of which is on a hexagonal ring.
Fig.~\ref{frustrated_structures} and Fig.~\ref{hexagonal_structures}
serve as keys to the labeling of the sites.

All of the nearest neighbor correlation functions are negative, which
is not surprising since the ground state wavefunction is chosen to
minimize the sum over these correlations.  In order to provide physical
insight into the results, we consider the following argument:  it is
possible to solve the HAFM analytically on a structure consisting of a central
site and its three neighbors.  The sum of the three correlations for this
system is $-5/4$.  The variational principle can then be used to show
that for a general structure, the sum of the three correlations between
a given site and its neighbors can not be less than $-5/4$.  This sum is
reduced in magnitude by frustration and by quantum fluctuations when
additional sites are included in the structure.  However, the existence
of the strict bound discussed above
suggests that a strong correlation between a site and one of its
neighbors will reduce the correlations to the rest of its neighbors.
This behavior is exemplified by the correlations in Table \ref{fs_results}.
The strongest correlations, those in the $H-H$ column, are for the
bonds between two sites that are on the same hexagonal ring.
Furthermore, the strongest of these correlations are found on the
frustrated structures where the bonds that form the hexagonal ring do
not have to compete with two other identical bonds.  The drop in energy
between F-28 and F-32 can be attributed to an increase in the number of
bonds of this type.  The weakest nearest neighbor correlations are
found between sites that are located on different hexagonal rings.
These bonds are frustrated and also suffer from strong competition from
the bonds on each of the hexagonal rings.  To illustrate these arguments
in a specific example, consider the
F-26 structure.  The $C_{1,9}$ and $C_{11,12}$ correlations are both
frustrated since each of these bonds is included in two pentagonal
rings.  The $C_{1,9}$ correlation is much weaker (-0.103)
than the $C_{11,12}$ correlation (-0.332) because the $C_{1,9}$
correlation has competition from four strong (-0.424) correlations
of the $C_{1,2}$ type (correlations between sites that are on
the same hexagon but not on any other hexagons).  For similar
reasons, the $H-P$ correlations are weaker than the $P-P$
correlations.

The correlation functions for the 28 site frustrated structure are
constrained by the symmetries of the wavefunction, and this results
in several anomalously small correlations, especially $C_{3,13}$ for the
$A$ wavefunction.  Although the original structure is tetrahedral,
the process of resolving the two degenerate states breaks this symmetry
by singling out the symmetry axis through the bond between sites
$19$ and $20$.  There is an approximate equivalence of correlations
between the results for the two wavefunctions.  The role of $C_{1,2}$
is switched with $C_{2,3}$, the role of $C_{1,9}$ is switched with
$C_{3,13}$, and the role of $C_{2,11}$ is switched with $C_{6,7}$.
Roughly speaking, the correlations that are closest to the axis through
the bond between sites $19$ and $20$ switch places with the
correlations that are furthest away form this axis.  The F-28 structure
has unusually strong long range correlations between the sites labeled
as 7, 11, 15, and 28 in Fig.~\ref{frustrated_structures}~(d).  These
sites form the corners of a tetrahedron.  For the F-28A state, the
correlations of this type perpendicular to the symmetry breaking axis
($C_{7,28}$ and $C_{11,15}$) are $0.141$ and the other correlations of
this type ($C_{7,11}$, $C_{7,15}$, $C_{11,28}$, and $C_{15,28}$) are $0.136$.
For the F-28B state, these correlations are $0.134$ and $0.139$ respectively.
This result is interesting because it suggests strong ferromagnetic
correlations between the spins on the four apex sites that form the corners of
a tetrahedron in F-28.  This is consistent with quantum mechanical
calculations of the electronic structure of the C$_{28}$ molecule,
which is believed to have the same structure as the F-28 cluster:
in those calculations, the molecule is found to have an $s = 2$
ground state, with the spins in the four apex sites aligned~\cite{pederson93}.

\section{Conclusion}
\label{Conclusion}

The variational Hilbert space truncation approach provides an effective
way to extend the range of structures for which exact diagonalization of
the HAFM is feasible.  Substantial reductions in memory can be obtained
with less than a 1\% error in the ground state energy.  A few
percent error is introduced in most correlations.  The exception is
very weak correlations for which the method will give a rough idea
at best.  For system sizes that are at the current leading edge of
computational capabilities, a reduction of the Hilbert space by a
factor of thirty can be achieved.  For the HAFM, a factor of thirty
reduction in memory use allows structures with about $5$ additional
sites to be handled.  Our method is compatible with symmetrization
techniques, which, depending on the structure under consideration,
can achieve a similar reduction in memory requirements.  Finally,
our method should be useful for models other than the HAFM.
In fact, much larger reductions in the size of the Hilbert space can
be expected for systems where the ground state is dominated by a few
of the basis states used in the expansion of the wavefunction.
For such systems, the method should be capable of identifying the
important basis states, and thus the important physics of the ground state.

Using this variational approach, we have successfully determined the
ground state properties of the
HAFM on a series of frustrated and unfrustrated structures.  An
interesting and unexpected result is the doublet nature of the
ground state of the 28 site frustrated structure.  The 32 site
frustrated structure seems to be a rotational singlet, but it
would be interesting to know whether other larger structures of
this type also break structural symmetries.

\section*{Acknowledgement}

This work was supported by ONR Contract \#N00014-93-1-0190.
The computations were performed on the NRL 256-node CM-5 supercomputer.
We acknowledge helpful input during the initial stages of this project
from Prof. L. Johnsson.

\appendix
\section*{Implementation of the Method in a Massively Parallel Architecture}
\label{Implementation}

The largest size truncation that we solved consisted of 20 million
states, which is 3.33\% of the full-space of the F-32 structure.
The diagonalization of such matrices is a time consuming process.
Our implementation on the CM5 massively parallel architecture
provided a vivid demonstration of the conflict between efficient use of
memory and efficient use of CPU time.  The Hamiltonian matrix
can either be stored in core memory or generated during each matrix-vector
multiply required by the Lanczos method.  Storing the Hamiltonian reduces
the time by about a factor of ten at the expense of a factor of
four increase in the memory.  A third possibility would be to store
the Hamiltonian on an external device with fast access, such as the
Scalable Disk Array (SDA).  Because the SDA's total capacity is only
about three times that of the core memory, we have not implemented
this option.

Multiplication of the wavefunction by the unstructured, sparse Hamiltonian
matrix requires general communication between sections of memory distributed
to different processors, and therefore it is not expected to parallelize
efficiently.  Such multiplications form the core of the Lanczos algorithm.
Careful implementation of these multiplications as well as the generation
of the new truncations and Hamiltonians is essential to good parallel
performance.  We separate the techniques used to obtain reasonable
efficiency, while avoiding excessive memory use, into three categories:
the use of previous results during the generation of new results, the
balanced division of work over both processors and time (load balancing),
and the usage of sorting instead of searching.  These are discussed in
order:

{\it 1. Use of previous results:}
There are three tasks that must be accomplished during each iteration
of the variational Hilbert space truncation method:  generation of the
truncated space that will be investigated during the iteration,
generation of the corresponding Hamiltonian, and diagonalization of the
Hamiltonian.  Since each truncation is a variation of the previous
truncation, it is possible to use results from the previous
iteration to speed up the calculation considerably.  The most important
gain in efficiency is obtained by initializing the Lanczos routine
with a guess wavefunction derived from the results for the previous
iteration by using first order perturbation theory.  This requires
very little extra work since all of the expensive steps of the
perturbation theory are already carried out as part of the generation
of each new truncation.  This procedure can reduce the time required
to find the ground state by a factor of $100$.

{\it 2. Load balancing:}
In order to get a reasonable rate of performance out of a parallel
computer, it is necessary to group sets of similar operations together.
On the other hand, avoiding the use
of large amounts of memory requires divying up similar operations
over time so that the memory needed to perform each group of operations
can be reused.  Therefore, getting good utilization of both processors
and memory requires groups of operations that are neither too big,
nor too small.  In general, the best performance is achieved
by identifying the largest unavoidable use of memory, and then using
groups of operations that are somewhat smaller.  One example is the
generation of the Hamiltonian, where the best compromise is to consider
all of the elements resulting from exchanging one pair of nearest neighbor
spins at the same time.  Another example is provided by the selection of each
new truncation.  It is necessary to compute $\nabla_{\beta}E$ for each
trial state $|\beta\rangle$
[see Eqs. (\ref{inside_force}) and (\ref{outside_force})].
For most cases of interest, there are
many more trial states than states in the truncation.  Thus, if the
Hamiltonian is not stored, it is efficient in terms of memory usage to divide
the trial states into smaller sets and compute for one set at a time.
This is possible since only the $N_{trunc}$ states with the largest
values of $\nabla_{\beta}E$ need to be retained at each step.  This
technique allows sets of trial states of arbitrary size to be considered
when generating each variational step.

{\it 3. Sorts instead of searches:}
A standard problem encountered during numerical calculations involving
spin models is that information (in this case, the components of
the wavefunction) about each of the spin configurations must be packed
into memory in some manner that allows its quick retrieval.  It is
trivial to associate each spin configuration with a unique number,
but the resulting set of numbers is not usually dense.  Considerable
research effort has been expended in developing efficient hashing
routines for locating the memory addresses associated with a given
spin configuration~\cite{gagliano86,lin90}.  Since the set of basis
states changes stochastically during each iteration of our variational
truncation process, the algorithm requires a flexible hashing procedure
without substantial overhead for setup.  For our implementation, we also
needed a procedure that parallelizes efficiently.  The radix sort
algorithm, which consists of hashing on successive blocks of bits,
sorts a list of $N_{keys}$ in a time proportional
to $N_{keys}$.  This algorithm parallelizes ideally (it uses a time
proportional to $N_{keys}/N_{proc}$ on a machine with $N_{proc}$
processors) and is stable (if two entries are equal, the entry with
the smaller initial subscript will be sorted to the location with
the smaller final subscript).  This points to combining many searching
operations together and using sorts to do searching efficiently on a
parallel machine.  We implemented a procedure based on inter-sorting a list
of states with unknown memory addresses with a list of all the states
in the truncation.  This procedure worked so well that we were able
to generate the Hamiltonian during each matrix-vector multiply rather
than storing it, thereby saving on memory usage and extending the size of
the system that could be handled.

\begin{figure}
\caption{(a) 20-site, (b) 24-site, (c) 26-site, (d) 28-site, (e) 32-site,
and (f) 60-site fullerene-like frustrated structures with spherical topology.
The 3-dimensional structures are shown projected on a plane, which introduces
a distortion of relative distances.  Therefore, the figures only indicate
the connectivity of the the structures, and the apparent lengths of the
bonds are not to be interpreted literally.}
\label{frustrated_structures}
\end{figure}

\begin{figure}
\caption{(a) 18-site, (b) 24-site, and (c) 26-site unfrustrated structures
derived from the honeycomb lattice by applying periodic boundary conditions
along the thinner solid lines.}
\label{hexagonal_structures}
\end{figure}

\begin{figure}
\caption{Error in the ground state energy resulting from truncation of the
Hilbert space for (a) some frustrated structures and (b) some unfrustrated
structures.  Lines indicate the results of truncating the Hilbert space
based on the weights of states in the full
space solution.  Individual points indicate results from the variational
method and correspond to the same structure as the line immediately above them.
The stars indicate results for the 28 site system, where results from
truncating based on the full-space solution are unavailable.}
\label{energy_errors}
\end{figure}

\begin{figure}
\caption{Spin-spin correlations for the unfrustrated structures based on the
honeycomb lattice as a function of the fraction of states retained in the
truncated Hilbert space.  Lines indicate results from truncating the Hilbert
space based on the weights of states in the full-space solution.
Individual points indicate results from the
variational method.  The groups of lines are labeled by nearest neighbor
distances. The points near zero abscissa correspond to small, but finite,
truncations.}
\label{best_case_corr}
\end{figure}

\begin{figure}
\caption{Fractional error in the smallest spin-spin correlations on the
frustrated structures as a function of the fraction of states retained in
truncated Hilbert space.  Lines indicate results from truncating the
Hilbert space based on the weights of states in the full-space solution.
Individual points indicate results from the variational method.
}
\label{worst_case_corr}
\end{figure}

\begin{table}
\caption{Some properties of the structures considered in this paper.}
\label{struct_props}
\begin{tabular} {|c|c|c|c|c|}
\hline
\multicolumn{5}{|c|}{\bf Structural Properties} \\
\hline
\bf Structure & \bf Point Group Symmetry & \bf Pentagons & \bf Hexagons &
\bf Dimension of Hilbert Space\\
\hline
\hline
F-20  & $I_{h}$              & 12  &  0  &     184,756 \\
\hline
F-24  &  $D_{6d}$            & 12  &  2  &   2,704,156 \\
\hline
F-26  &  $C_{3v}$            & 12     &  3  &  10,400,600 \\
\hline
F-28  &  $T_{d}$             & 12  &  4  &  40,116,600 \\
\hline
F-30  &  $C_{2v}$            & 12  &  5  &  155,117,520 \\
\hline
F-32  &  $D_{3}$             & 12  &  6  & 601,080,390 \\
\hline
F-60  &  $I_{h}$             & 12  &  20  & $1.2 \times 10^{17}$ \\
\hline
\hline
H-18  &  $C_{3v}$            &  0  &  9  &       48,620 \\
\hline
H-24  &  $C_{3v}$            &  0  & 12  &   2,704,156 \\
\hline
H-26  &  $C_{3v}$            &  0  & 13  &  10,400,600 \\
\hline
\end{tabular}
\end{table}

\begin{table}
\caption{Ground state energy and nearest neighbor correlation
  functions of each structure.  Results for the F-32 structure are
  from a truncation retaining 20 million of the 601 million states.
  All other results are from full-space solutions.}
\label{fs_results}
\begin{tabular} {|c|c|c|c|c|c|}
\hline
\multicolumn{6}{|c|}{\bf Ground State Properties} \\
\hline
\bf Structure &
   \bf $\hbox{E}_{0} / \hbox{Site}$ &
    \multicolumn{1} {c|}{\bf $H-H$} &
    \multicolumn{1} {c|}{\bf $H-H^{\prime}$} &
    \multicolumn{1} {c|}{\bf $H-P$} &
    \multicolumn{1} {c|}{\bf $P-P$} \\
\hline
\hline
F-20  & $-1.722219$  &                       &                       &
                            & $C_{1,2} = -0.324$ \\
\hline
F-24  & $-1.726614$  &  $C_{1,2} = -0.409$   &                       &
        $C_{1,9} = -0.203$  & $C_{8,9} = -0.371$ \\
\hline
F-26  & $-1.719921$  &  $C_{1,2} = -0.424$   &  $C_{1,9} = -0.103$   &
        $C_{2,11} = -0.265$  & $C_{11,12} = -0.332$ \\
      &              &  $C_{2,3} = -0.339$  &                       &
                            &                    \\
\hline
F-28A & $-1.719633$  &  $C_{1,2} = -0.275$   &  $C_{1,9} = -0.327$   &
        $C_{2,11} = -0.321$  &                   \\
      &              &  $C_{1,6} = -0.362$   &  $C_{3,13} = -0.063$  &
        $C_{6,7} = -0.269$   &                   \\
      &              &  $C_{2,3} = -0.425$   &                       &
                             &                   \\
\hline
F-28B & $-1.719633$  &  $C_{1,2} = -0.433$   &  $C_{1,9} = -0.151$   &
        $C_{2,11} = -0.286$  &                   \\
      &              &  $C_{1,6} = -0.346$   &  $C_{3,13} = -0.415$  &
        $C_{6,7} =  -0.338$  &                   \\
      &              &  $C_{2,3} = -0.283$   &                       &
                             &                   \\
\hline
F-32  & $-1.736$     &  $C_{2,3} = -0.420$   &  $C_{3,4} = -0.101$   &
        $C_{1,2} =  -0.279$  &  $C_{12,13} = -0.333$ \\
      &              &  $C_{2,10} = -0.352$  &  $C_{11,22} = -0.123$ &
                             &                       \\
      &              &  $C_{3,13} = -0.407$  &                       &
                             &                       \\
      &              &  $C_{4,15} = -0.509$  &                       &
                             &                       \\
      &              &  $C_{11,12} = -0.335$ &                       &
                             &                       \\
\hline
H-18  & $-1.871907$  &  $C_{1,2} = -0.374$   &                       &
                             &                   \\
\hline
H-24  & $-1.860839$  &  $C_{1,2} = -0.370$   &                       &
                             &                   \\
\hline
H-26  & $ -1.858385$ &  $C_{1,2} = -0.369$   &                       &
                             &                   \\
\hline
\end{tabular}
\end{table}


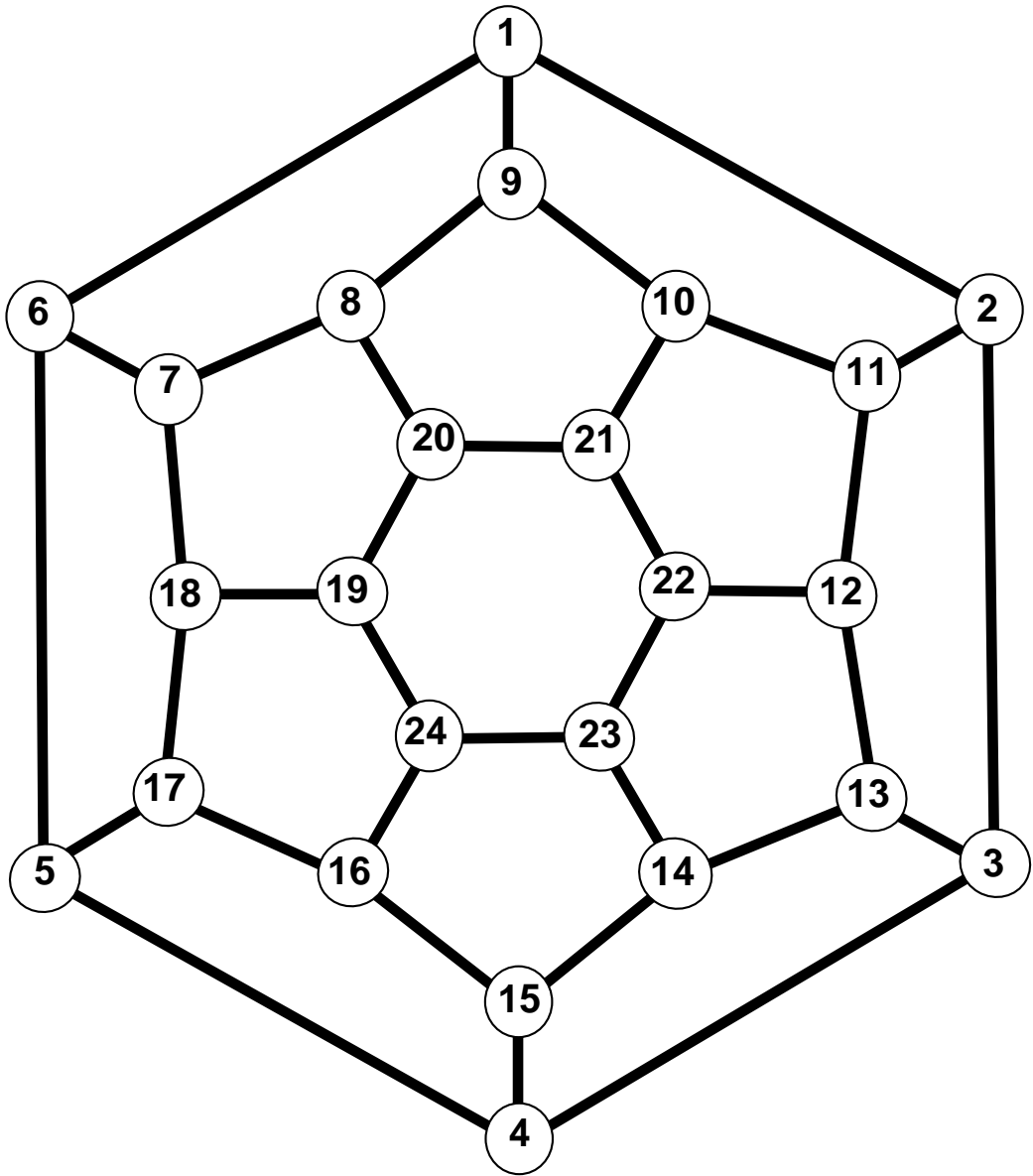
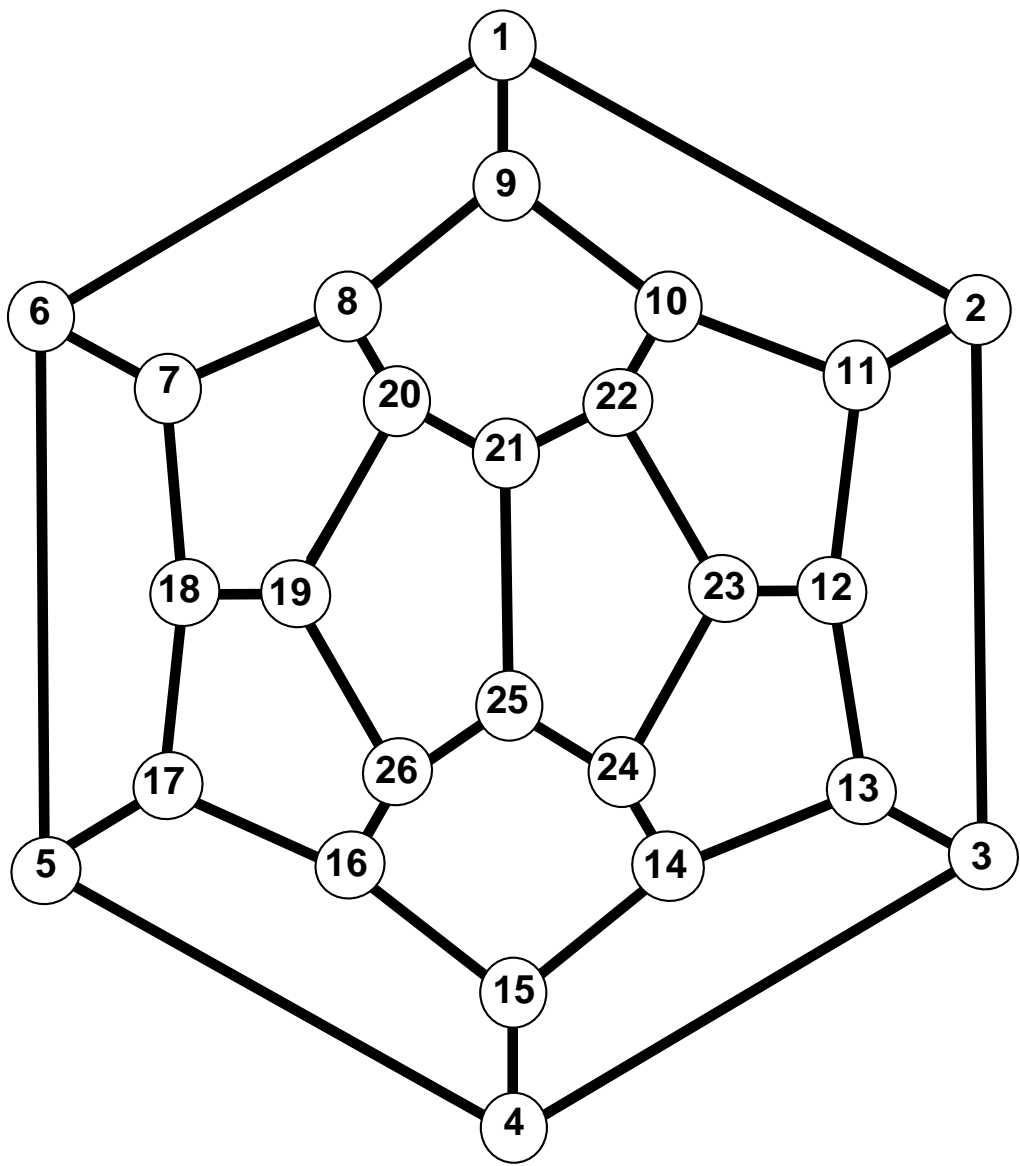
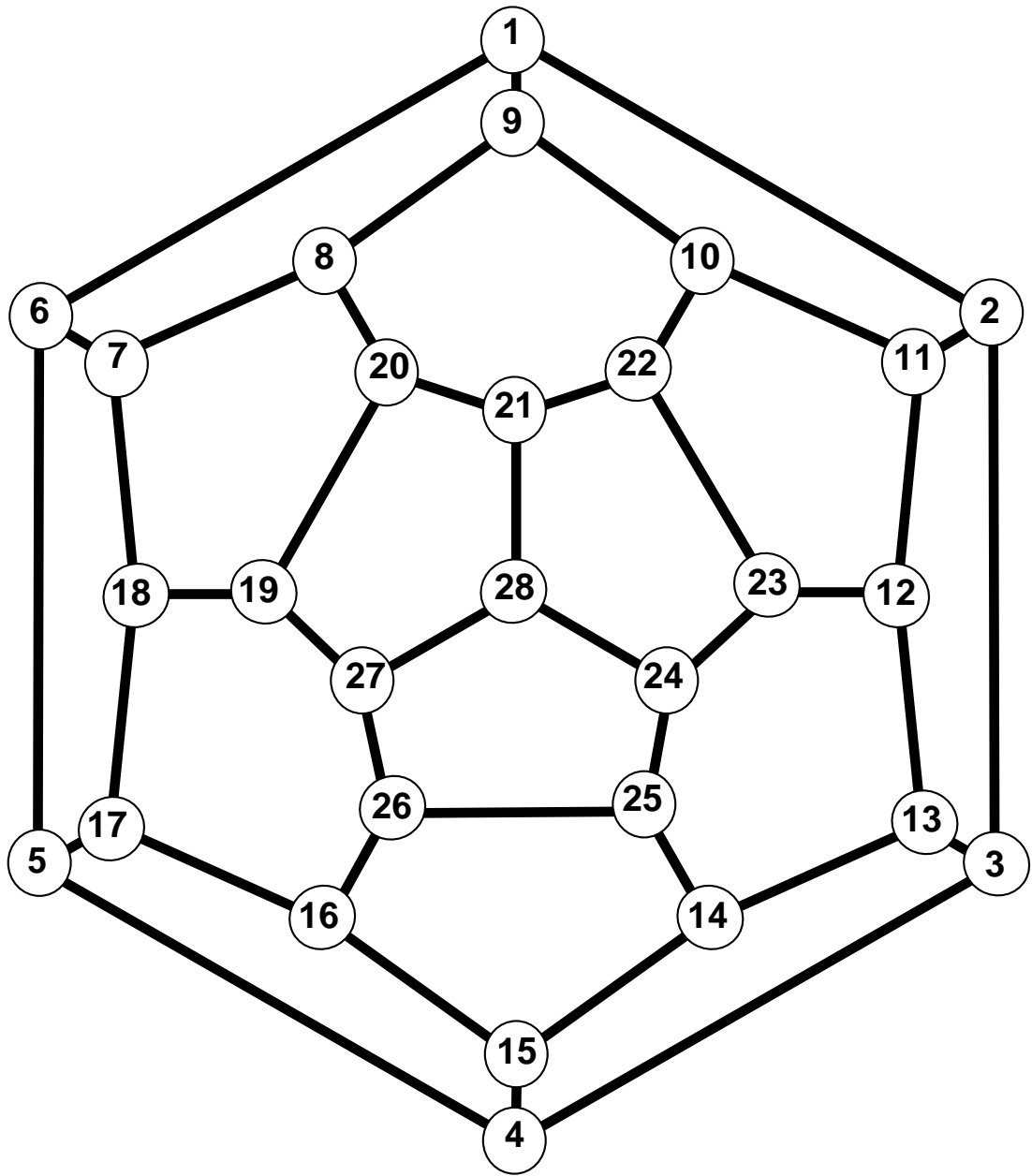
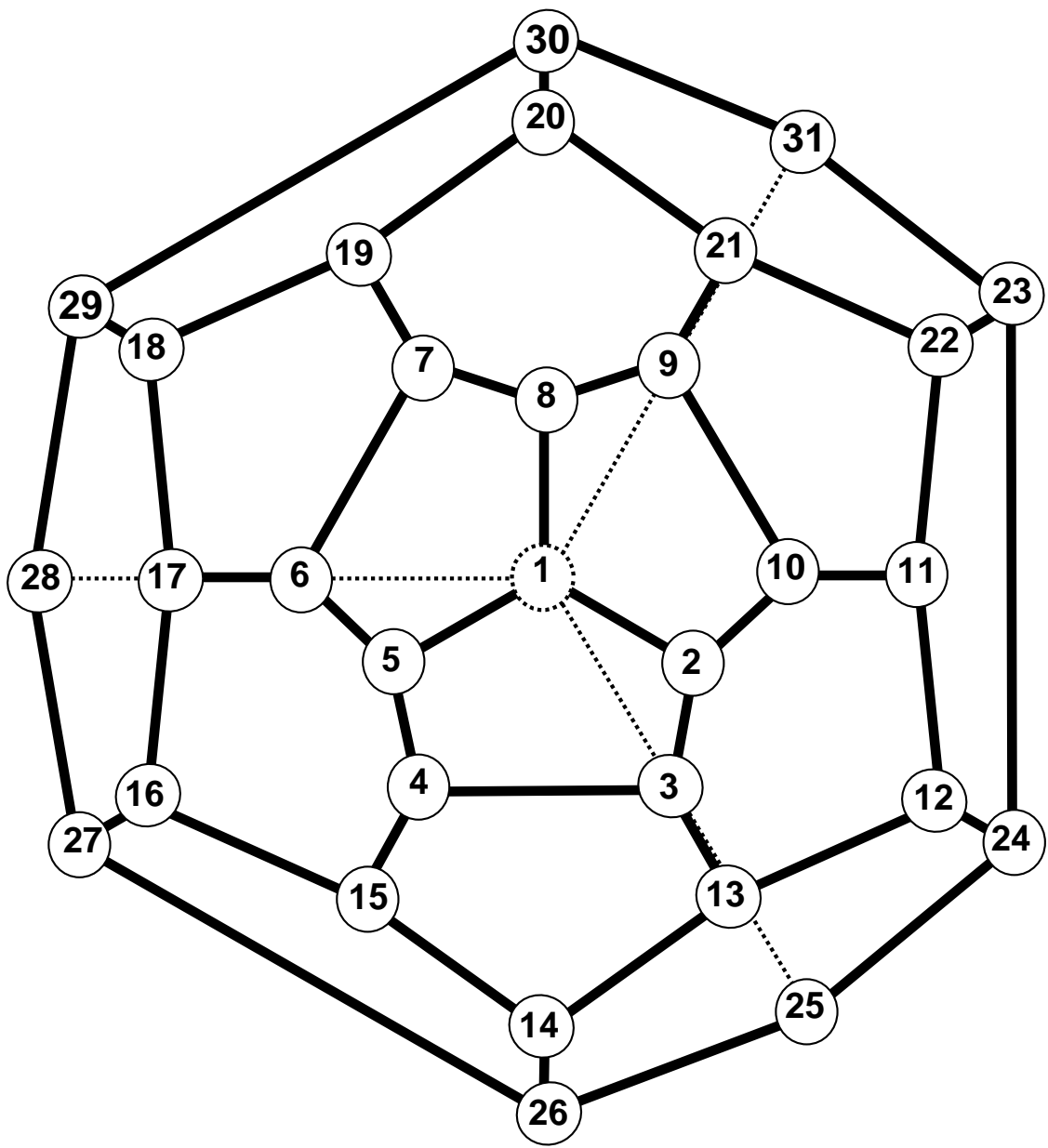
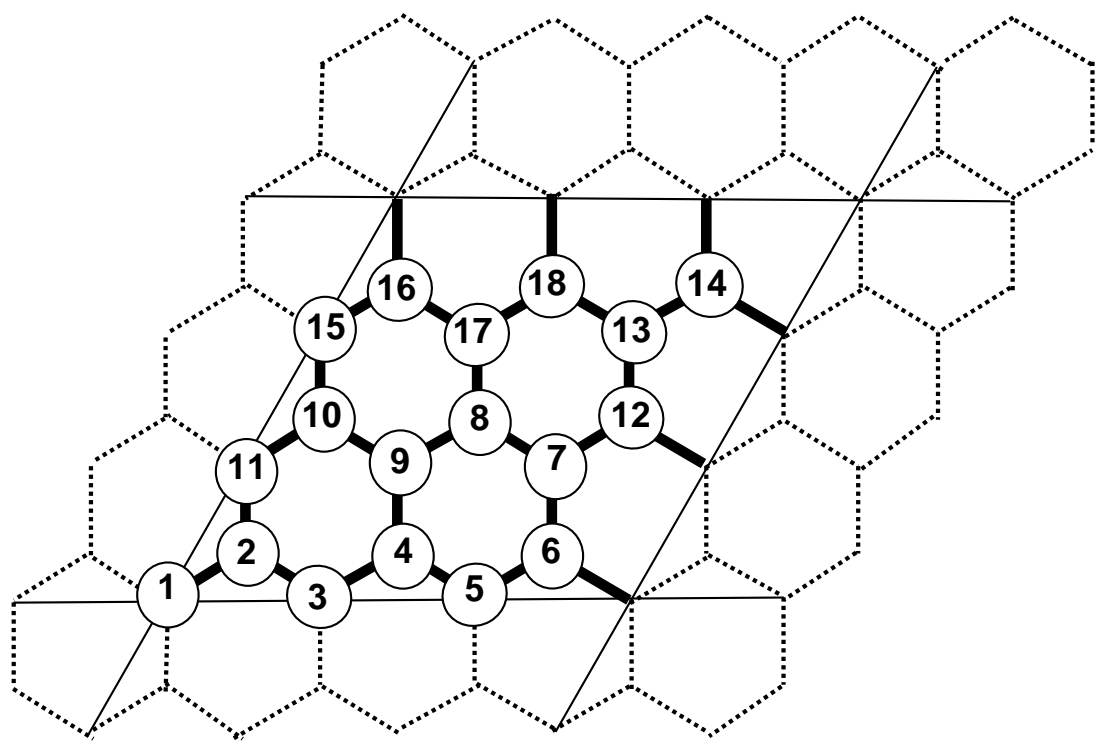
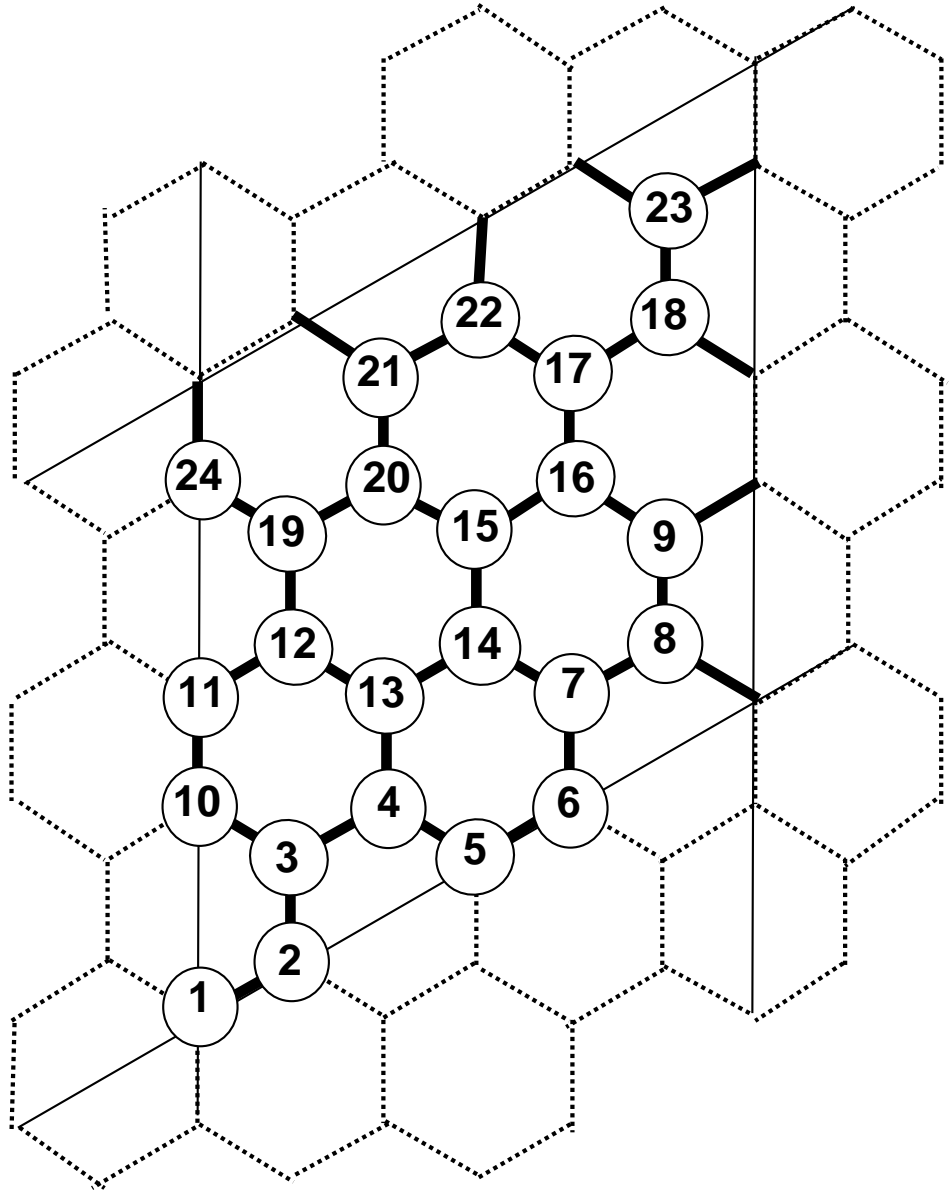
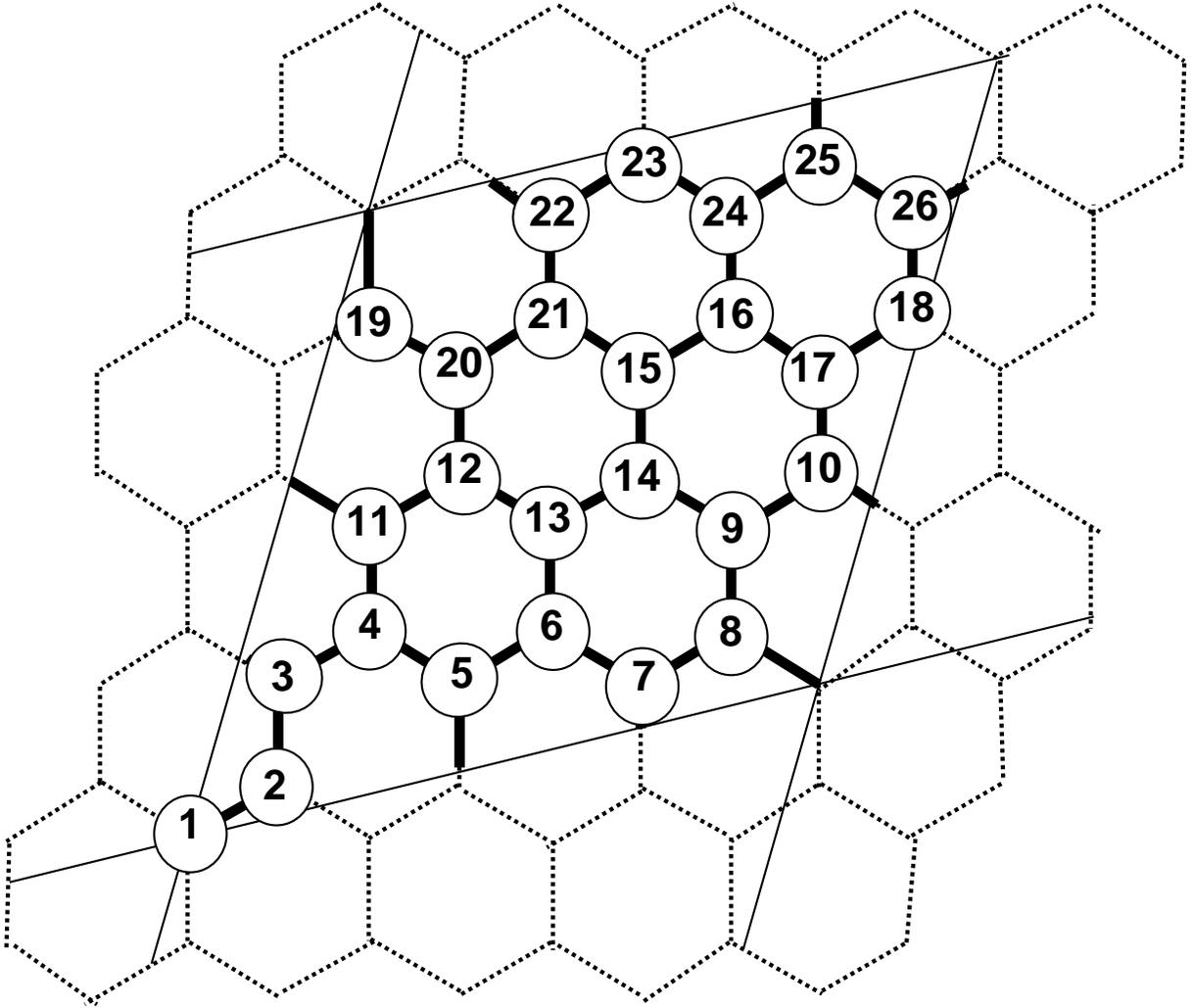
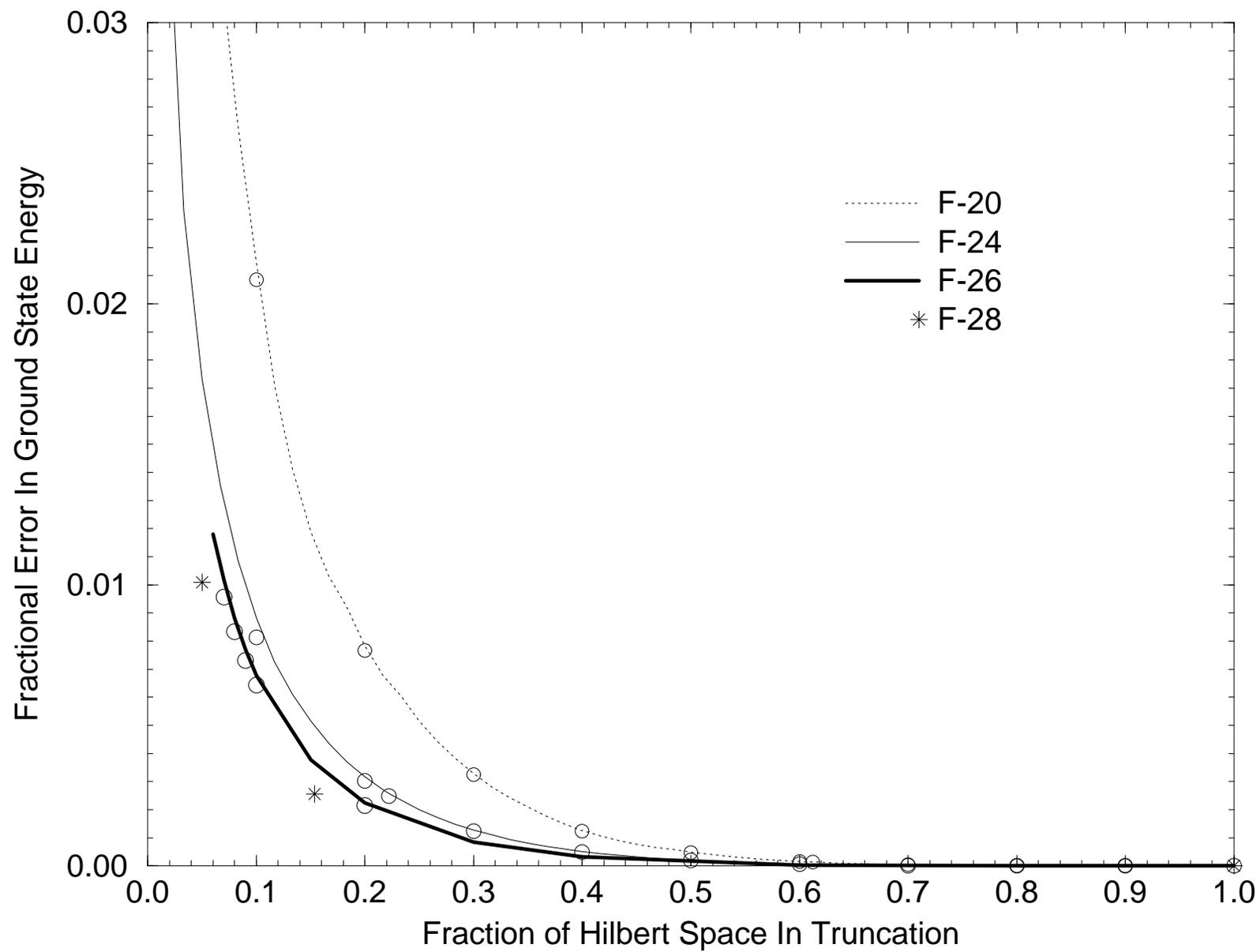
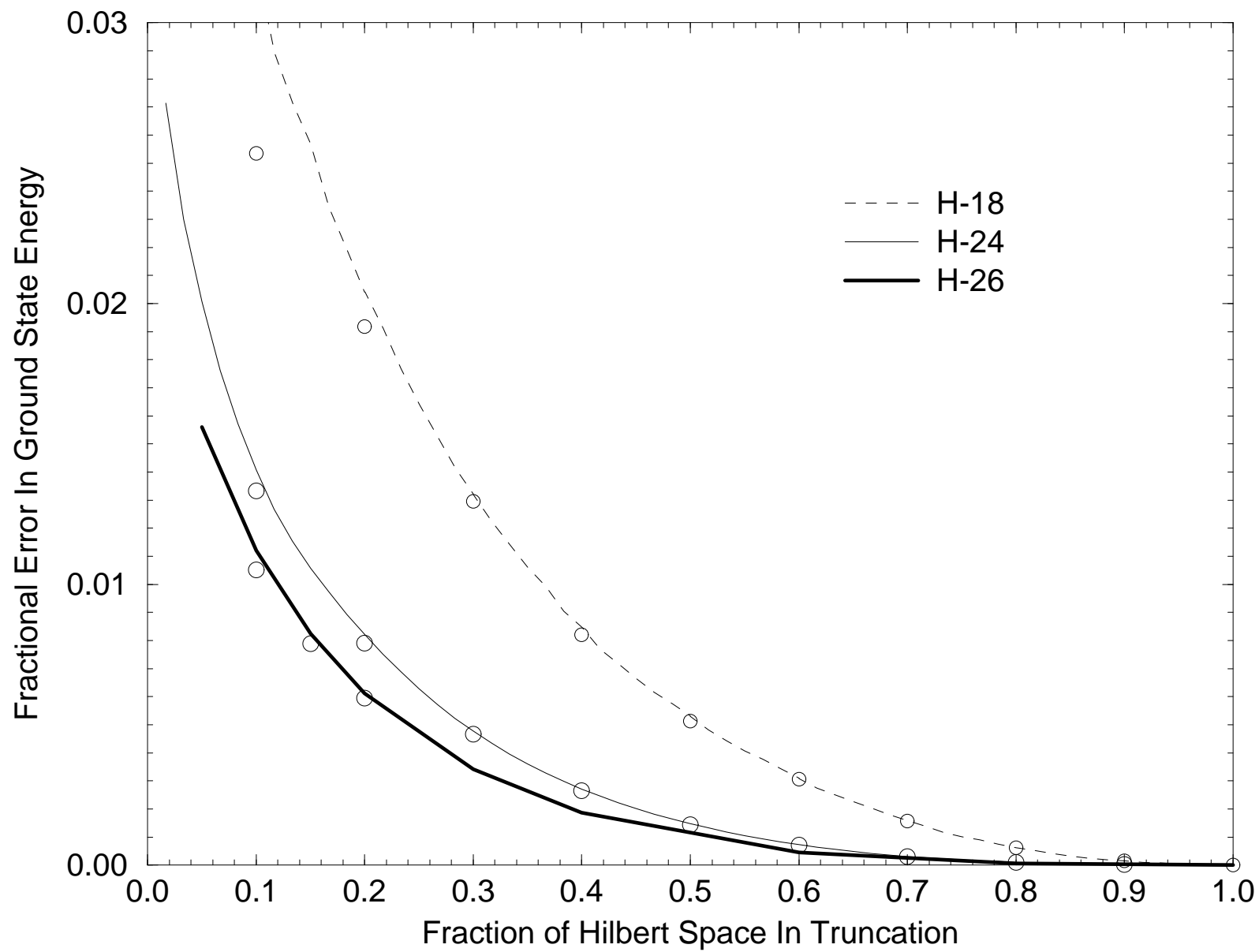
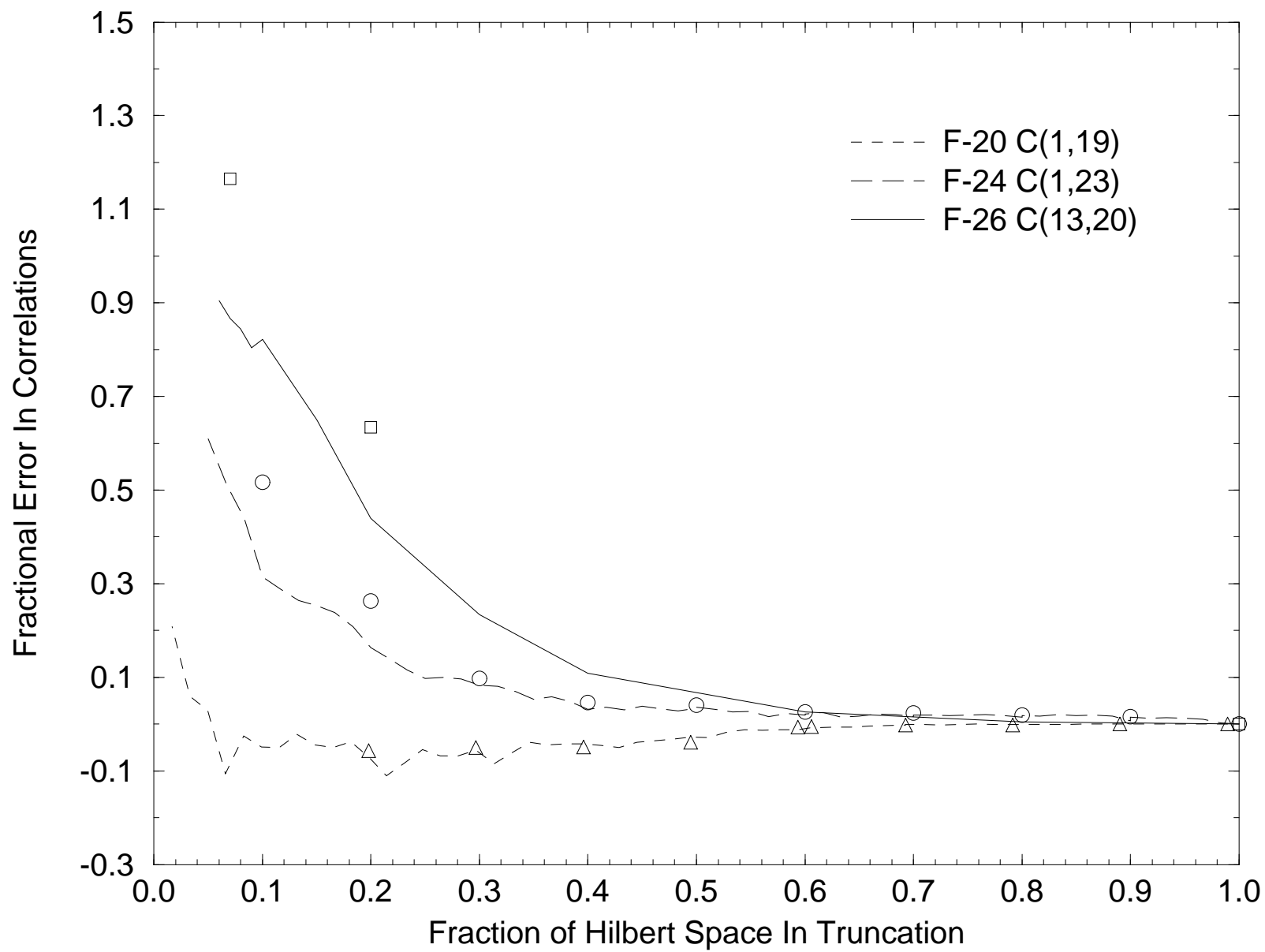
\end{document}